\documentclass[10pt]{article}
\pdfoutput=1

\newlength{\abstractwidth}
\usepackage{booktabs}
\usepackage{array}
\usepackage{amsmath}
\usepackage{amsfonts}
\usepackage{amssymb}
\usepackage{latexsym}
\usepackage{appendix}
\usepackage{amsthm}
\usepackage{mathtools}
\usepackage{physics}

\numberwithin{equation}{section}

\usepackage{bbold}

\usepackage{epsf}
\usepackage{color}
\usepackage{graphicx}
\usepackage{dsfont}
\usepackage{subfigure}

\usepackage[export]{adjustbox}
\usepackage{hyperref}
\definecolor{darkred}{rgb}{0.8,0.1,0.1}
\hypersetup{colorlinks=true, linkcolor=darkred, citecolor=blue, linktoc=page}

\usepackage{caption}
\usepackage{subcaption}

\graphicspath{{Figures/}}

\newcommand{\be}{\begin{equation}}
\newcommand{\ee}{\end{equation}}

\newcommand{\<}{\langle}
\renewcommand{\>}{\rangle}

\def\bsk{{\boldsymbol{k}}}

\def\bsq{{\boldsymbol{q}}}

\def\bsx{{\boldsymbol{x}}}

\def\bsy{{\boldsymbol{y}}}

\def\bsk{{\boldsymbol{k}}}
\def\bsq{{\boldsymbol{q}}}

\def\Im{{\rm Im \,}}

\def\p{\partial}

\def\eps{\epsilon}

\def\m{\mu}
\def\n{\nu}
\def\){\right)}
\def\({\left( }
\def\]{\right] }
\def\[{\left[ }

\def\no{\nonumber}

\definecolor{ascolor}{rgb}{1,0,1}
\usepackage[normalem]{ulem}

\DeclareRobustCommand\asout{\marginpar{\color{ascolor}$\spadesuit$}\bgroup\markoverwith{\color{ascolor}{\rule[0.4ex]{2pt}{0.8pt}}}\ULon}

\makeatother

\begin{document}


\begin{center}
 {\Large \bf  Squeezed-state radiation in shockwave scattering:\\
  QCD-Gravity double copy}
 \vskip 0.4in
 {\large   Anna M. Sta\'sto$^a$, Himanshu Raj$^b$, Raju Venugopalan$^{b,c,d}$} 
 \vskip .2in

$^a$ {\it Department of Physics, Penn State University, University Park, PA 16802, USA}\\[0.5cm]

 $^b$ {\it Center for Frontiers in Nuclear Science, Department of Physics and Astronomy,}\\
 {\it Stony Brook University, Stony Brook, NY 11794, USA}\\[0.5cm]

$^c${\it  Department of Physics, Brookhaven National Laboratory,}
 {\it  Upton, NY 11973, USA} \\[0.5cm]
 
 $^d${\it Higgs Center of Theoretical Physics, The University of Edinburgh,}
 {\it Edinburgh, UK}
 \vskip 0.1in
 \begin{abstract}
Gluon and graviton radiation in strong field shockwave scattering are described by effective Lipatov vertices, with the graviton Lipatov vertex  proportional to the bilinear of its QCD counterpart. We show here that the n-particle gluon radiation spectrum can be described as a generalized Susskind-Glogower (gSG) squeezed coherent state and discuss the properties of such squeezed states. The double copy structure of the radiative frameworks suggests that multi-graviton radiation can be similarly described as a gSG state. We examine the physical parameter space and show that very large squeezing parameters $\sim \ln({\bar n})$ (where ${\bar n}$ is the mean graviton occupancy) are feasible for nearly minimal uncertainty configurations of the gSG state. Quantum noise in the corresponding gravitational wave spectrum is enhanced above the sensitivity of current and future gravitational wave detectors. Our results point to the importance of a comprehensive study of the strong field Lipatov regime of gravitational radiation.  
 \end{abstract}
 \end{center}

\baselineskip=16pt
\setcounter{equation}{0}
\setcounter{footnote}{0}

\section{Introduction}
A question of great interest in QCD is the many-body structure of quark-gluon matter at ultrarelativistic energies. Since quarks and gluons are not observable, this information has to be extracted by looking for its imprint on hadronic final states. 
A highly influential body of work in this direction were the papers by Bialas and Peschanski (almost exactly forty years to date) who proposed studying the moments of multiplicity distributions to look for self-similar structure on different scales~\cite{Bialas:1985jb,Bialas:1988wc}. In particular, their suggestion to focus on the rapidity moments of high multiplicity events was well ahead of its time and anticipated the explosion of interest in the ``ridge-like" structure of such events following their discovery at RHIC and the LHC~\cite{Dusling:2015gta}. It is therefore especially fitting to contribute this paper on some remarkable features of multi-particle production in QCD and in gravity to this volume in honor of Prof. Bialas' 90th birthday. 

Our goal here is to explore the possibility that gravitational radiation from the  $2\rightarrow n$ emission of gravitons in ultrarelativistic black hole collisions is in the form of a squeezed state.\footnote{For alternative discussion of the possibility of squeezed state radiation in binary black hole collisions, see for instance \cite{Kanno:2025how,Kanno:2025fpz,Das:2025kyn}.} Specifically, our focus will be on the copious radiation emitted when the impact parameter of scattering approaches the Schwarzchild radius of the black hole. The argument underlying this idea is motivated by the the observation that inclusive graviton radiation amplitude in shockwave scattering is a double copy of  inclusive gluon radiation from shockwave scattering in high energy QCD~\cite{Raj:2023iqn}. In QCD, analytical solutions of Yang-Mills equations in shockwave scattering (within so-called dilute-dilute and dilute-dense approximations) reveal that gluon radiation is described by the Lipatov vertex~\cite{Blaizot:2004wu,Gelis:2005pt}, a building block of $2\rightarrow n$ scattering amplitudes in the Regge asymptotics of QCD~\cite{Kuraev:1976ge,Balitsky:1978ic}. Likewise, solutions of Einstein's equations for the scattering of Aichelburg-Sexl shockwaves in an analogous dilute-dilute\footnote{While the bilinear of the QCD Lipatov vertex contributes to dilute-dense shockwave solutions of Einstein's equations in gravity, the result in this case is not a simple double copy~\cite{FRV}.} expansion~\cite{Raj:2023iqn,Raj:2023irr} are described by a gravitational Lipatov vertex~\cite{Lipatov:1982it,Lipatov:1982vv}, which is proportional to a  bilinear of the QCD Lipatov vertex. We will show here that QCD radiation in shockwave scattering can be described as a particular generalization of a single mode squeezed state. The identical double copy structure of $2\rightarrow n$ gravitational radiation in shockwave scattering is therefore strongly suggestive that it too can be described similarly. Whether that is indeed the case is a conjecture; we hope our results are sufficiently motivating to explore this line of inquiry further.

While the scattering of classical shockwaves is natural to consider in Einstein gravity, it is at face value deeply surprising in QCD, where one expects the physics to be fundamentally quantum. It is hence important at the outset to address how such a (semi-)classical picture arises in QCD before proceeding further with a discussion of this double copy as more than a mathematical curiosity. The relevant kinematics is the high energy multi-Regge asymptotics of perturbative QCD, where $2\rightarrow n$ multi-gluon radiation with increasing energy, is seen to be dominated by $s$-channel fractionation. It can be visualized as a vertical ladder structure where the emissions are strongly ordered in rapidity. The objects exchanged down the rungs of the ladder are ``reggeized" gluons, which arise from the resummation to all orders of the leading  virtual contributions in the Regge limit. The aforementioned Lipatov vertices are the effective vertices controlling the emission of  gluon from  rungs of the ladder that are ``cut", forming an ordered gluon cascade. 

This multi-gluon cascade is described by the BFKL equation~\cite{Kuraev:1976ge,Kuraev:1977fs,Balitsky:1978ic}, which leads to rapidly growing gluon distributions with increasing boost (rapidity). As phase space occupancies approach $1/\alpha_S$ (where $\alpha_S$ denotes the QCD coupling), screening and recombination effects qualitatively alter the gluon cascade, and generate instead a strongly correlated semi-classical state of overoccupied gluons. This phenomenon, called gluon saturation~\cite{Gribov:1983ivg,Mueller:1985wy}, is characterized by a 
dynamical ``close packing" scale $Q_S(x)\gg \Lambda_{\rm QCD}$ which evolves with the rapidity $Y= \ln(1/x)$, where $x$ is the fraction of the shockwave momentum carried by a ``wee" gluon in the cascade. 

The physics of the saturation regime of QCD is nonperturbative and is described by the Color Glass Condensate (CGC) semi-classical effective field theory~\cite{Iancu:2003xm,Gelis:2010nm}, where 
$\alpha_S (Q_S)\ll 1$, but the effective coupling $\alpha_S N \sim O(1)$, with occupancy $N\sim 1/\alpha_S$. The power counting in the CGC EFT is qualitatively different from perturbative QCD~\cite{McLerran:1993ka,McLerran:1993ni,McLerran:1994vd}. The dominant source of radiation in this regime arises from $t$-channel fractionation rather than the $s$-channel fractionation of perturbative QCD. It can in turn be visualized as a horizontal ladder consisting of several $t$-channel exchanges resulting from  coherent ``multiple scattering" of two semi-classical ``lumps" of overoccupied gluons. The rungs of this horizontal ladder are connected by gluons; cutting these, is what leads to gluon emission. 

The physical process most suited to this description is gluon radiation in collisions (at RHIC and the LHC) of two ultrarelativistic nuclei~\cite{Kovner:1995ja,Kovner:1995ts,Krasnitz:1998ns,Krasnitz:1999wc,Krasnitz:2000gz}. The large number of color charges that scatter coherently in these heavy-ion collisions naturally generate the conditions for overoccupancy. The Yang-Mills equations describe the copious radiation of gluons which form a far-from-equilibrium Glasma state~\cite{Lappi:2006fp,Gelis:2006dv}, whose subsequent evolution forms a thermal Quark-Gluon Plasma (QGP)-for a review of this thermalization process, see \cite{Berges:2020fwq}. 

 Analytical computations  of inclusive gluon production in the CGC EFT are feasible in the aforementioned dilute-dilute and dilute-dense approximations. For simplicity, we will restrict our discussion here to the dilute-dilute case. Since the correlation length of each semi-classical lump is $\sim 1/Q_S$, multi-gluon emissions occur in a ``flux tube" with a cross-section  $\sim 1/Q_S^2$.   The $n$-body distribution is obtained by sewing together multiple $2\rightarrow 3$ emissions, in the amplitude and complex conjugate amplitude (corresponding to cut gluons connecting the horizontal rungs of the ladder), where the emission vertex is the Lipatov vertex~\cite{Dumitru:2008wn}. The combinatorics of such gluon emissions results in the negative binomial   probability distribution (NBD) $P_{n;r}$. The parameter $r$ that characterizes the NBD is not arbitary but can be computed in the CGC EFT~\cite{Gelis:2009wh}; it is seen very simply to be proportional to the multiplicity of flux tubes, given by $R^2/(1/Q_S^2)= R^2\,Q_S^2$, where $R$ is the radius of the colliding nuclei (assumed to be identical). Note that NBDs have as limiting cases a Bose (or geometric) distribution for $r=1$ and a Poisson distribution for $r\rightarrow \infty$.

We will discuss here the $n$-gluon state corresponding to this negative binomial distribution and show that it corresponds to an ``intermediate phase"  coherent state~\cite{Fu:1996ra,Barnett01101998,Wang:1999mv}, belonging to the general class of  nonlinear coherent states~\cite{Manko:1996ppt}. As we will show, these have strong squeezing properties in specific parameter ranges. In the limiting case $r\rightarrow \infty$ it reduces to a coherent state where $P_{n;r}$, as noted, is a Poisson distribution. For $r=1$, the Bose (or geometric) distribution is a limiting case of a Susskind-Glogower type phase state~\cite{Susskind:1964zz}. We will call the NBD state for arbitrary $r$ the generalized Susskind-Glogower (gSG) state. 
Understanding the gSG structure of the Glasma may be important for a first principles understanding of how it thermalizes. Unlike photon squeezed states, gluons in this particular gSG state are strongly interacting, and may flow initially like a superfluid before decoherence sets in. See for instance \cite{Blaizot:2011xf,Berges:2019oun}. The description of superfluids as squeezed states has also been discussed extensively in cosmological contexts~\cite{Berezhiani:2025maf}.

In gravity, identically to the CGC EFT description, multi-graviton radiation arises from cutting gravitons linking rungs of horizontal ladders~\cite{Ciafaloni:2015xsr,Ciafaloni:2017ort}. As noted earlier,  this radiation is important only when impact parameters in the scattering approach the Schwarzchild radius from above: $b\rightarrow R_S$. In other words, it occurs in the strong field regime of the scattering. As in QCD, the building block of such multi-graviton emissions is the $2\rightarrow 3$ amplitude. In this case, the emission vertex is the gravitational Lipatov vertex, proportional, as noted, to the bilinear of its QCD counterpart. A key question is what the relevant scale is in sewing together the cut amplitudes in the amplitude and complex conjugate amplitude to generate the $n$-graviton state?  This is important to determine whether the radiation is in some form of squeezed state. Though we do not have a definitive answer, we will present arguments that it is very plausible that this state, like the CGC, is a gSG squeezed coherent state. 

The possibility that copious gravitational radiation can be emitted in a laser-like squeezed state is very interesting because quantum noise can be strongly enhanced in this radiation.  As is well-known, semi-classical effects demonstrating the underlying quantum nature of gravity are extremely difficult to observe empirically~\cite{Dyson:2013hbl}. This is due to the very large phase space occupancy of gravitons in gravitational waves. An interesting proposal by Parikh, Wilczek and Zaharaide~\cite{Parikh:2020nrd,Parikh:2020fhy,Parikh:2020kfh} (see also \cite{Guerreiro:2019vbq,Guerreiro:2021qgk, Abrahao:2023lle}) suggests that this difficulty may be overcome if the gravitational radiation is in the form of a squeezed state. On the surface, the root mean standard deviation in the arms of a gravitational wave detector due to quantum noise,  $\sigma_{\rm Planck}\propto L_{\rm Planck}\sim 10^{-35}$ m, which is what makes observing such effects challenging. However, if the radiation is in the form of a squeezed state, 
$\sigma_{\rm squeezed}\sim e^{\xi}\,\sigma_{\rm Planck}$, where $\xi$ is the squeezing parameter. If this quantity were O($30$),  quantum effects may be detectable if the measured gravitational radiation is in a squeezed state. The outstanding question therefore is under what conditions such strong squeezing is likely and we will outline some of the interesting possibilities in the context of gSG states. We will also briefly discuss the challenges involved in the measurement of squeezed states. 

The paper is organized as follows. In section II, we will  review the derivation of the $n$-gluon distribution in the dilute-dilute limit of shockwave scattering. 
We will discuss in Section III the corresponding generalized Susskind-Glogower squeezed state description of the Glasma. The gravitational double copy and the possible implications for the detection of quantum noise are discussed in Section IV. We conclude with a summary and discussion of extensions of the work presented here.  Some details of the derivation and a short primer on relevant general properties of coherent and squeezed states are provided in three appendices. 

\section{$n$-gluon distribution from shockwave scattering in QCD}

The arguments articulating how a shockwave picture of scattering in the Regge asymptotics of QCD arises have been reviewed elsewhere~\cite{Iancu:2003xm,Gelis:2010nm,Berges:2020fwq,Raj:2025hse}. 
The leading order contribution to multi-particle production in shockwave collisions is obtained from solutions to the Yang-Mills (YM) equations given by 
\begin{align}
\label{eq:YM-dense-dense}
D_\m F^{\m\n}=J^\n~,
\end{align}
where $F_{\m\n}=\p_\m A_\n-\p_\n A_\m+ig[A_\m,A_\n]$ is the field strength tensor and $J^\m$ is the covariantly conserved current, $ D_\m J^\m = 0$. For shockwave scattering of nuclei (heavy-ion collisions at ultrarelativistic energies), the shockwave currents can be represented as 
\be
J^{\n,a} = \delta^{\n+}\rho^a_L(x_\perp)\delta(x^-) + \delta^{\n-}\rho^a_H(x_\perp) \delta(x^+)
\,.
\ee
Here $\rho^a_L(x_\perp)$ and $\rho^a_H(x_\perp)$ are quasi-classical color charge distributions of each of the shockwaves (high energy nuclei). These color charges correspond to a higher dimensional representation of the color algebra resulting from the direct product of the large number of gluon color charges that are emitted as the nuclei are boosted~\cite{McLerran:1993ka,McLerran:1993ni,McLerran:1994vd,Jeon:2004rk}. For $A\gg 1$, the weight functional  $W[\rho_{L,H}]$ (corresponding to the distribution of color charge representations over the most likely classical representation) is  Gaussian distributed such that $\langle \rho_{L,H}^a(x_\perp)\rho_{L,H}^b(y_\perp)\rangle = Q_S^2\, \delta^{ab}\, \delta^{(2)}(x_\perp-y_\perp)$, with $Q_S^2 \propto A^{1/3}\, \Lambda_{\rm QCD}^2$. For simplicity, we will assume that the shockwaves/nuclei are identical. The $\delta(x^\mp)$  denote eikonal currents.  To leading order, they are independent of the light cone times $x^\pm$, respectively. 

Shockwave scattering in QCD \cite{Kovner:1995ja,Kovner:1995ts}, in general, can only be solved numerically~\cite{Krasnitz:1998ns,Krasnitz:2000gz,Berges:2020fwq} because the Yang-Mills fields are nonperturbatively large, $A_\mu\sim 1/g$. However, one can identify parameters $\rho_L/\nabla_\perp^2, \rho_H/\nabla_\perp^2$ in the YM equations that one can expand in to obtain analytic solutions. In CGC jargon, these are the dilute-dilute  $\rho_A/\nabla_\perp^2, \rho_B/\nabla_\perp^2 \ll 1$~\cite{Kovner:1995ja,Kovner:1995ts,Kovchegov:1997ke,Gyulassy:1997vt} and dilute-dense   $\rho_L/\nabla_\perp^2\ll 1, \rho_H/\nabla_\perp^2 \sim 1$~\cite{Dumitru:2001ux,Blaizot:2004wu,Gelis:2005pt} limits.  
The full nonperturbative solution is the dense-dense limit corresponding to $\rho_L/\nabla_\perp^2, \rho_H/\nabla_\perp^2 \sim 1$. 
\begin{figure}[ht]
\centering
\includegraphics[scale=1]{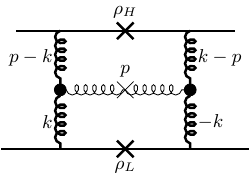} 
\caption{Dilute-dilute regime of shockwave scattering in QCD. Inclusive gluon radiation is depicted in light font by the emission of a gluon (where the crosses represent cuts putting the particles on-shell) in the amplitude and the complex conjugate amplitude. Color charge densities of the shockwave sources are denoted by $\rho_L$ and $\rho_H$. Reggeized gluons (in bold font) corresponding to classical fields generated by the sources interact via the effective Lipatov vertex (indicated by the black dot) to generate gluon radiation.}
\label{dilute-dilute}
\end{figure}

\begin{figure}[ht]
\centering
\includegraphics[scale=1]{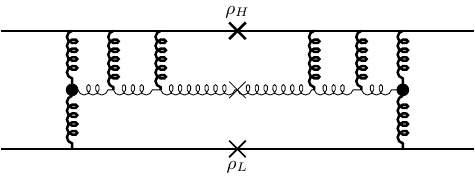}
\caption{Dilute-dense scattering of shockwaves, where $\rho_L/\nabla_\perp^2 \ll 1$ and $\rho_H/\nabla_\perp^2\sim 1$. For $\rho_H/\nabla_\perp^2\sim 1$, coherent multiple scatterings from the heavy shockwave can be resummed into a lightlike Wilson line. }
\label{dilute-dense}
\end{figure}

The dilute-dilute  case is illustrated in Fig.~\ref{dilute-dilute} and the dilute-dense case in Fig.~\ref{dilute-dense}. In the former, since $\rho_L/\nabla_\perp^2, \rho_H/\nabla_\perp^2 \ll 1$, coherent multiple scattering is suppressed in both of the colliding nuclei.  In the dilute-dense case,  multiple scattering insertions on the emitted gluon (from the dense color sources) are absorbed into a Wilson line. The gluon shockwave with $\rho_H(\bsx)$, at transverse position $\bsx$, moving in the positive $z$ direction is generated by the static current
\begin{equation}
J_{\mu}=g \delta_{\mu-} \delta\left(x^{-}\right) \rho_H(\bsx) ~,
\end{equation}
is 
\be
\label{eq:gluon-ShockBgnd}
\bar{A}_\m(x^-,\bsx) = -g \delta_{\mu-} \delta\left(x^{-}\right) \frac{\rho_{H}\left(\boldsymbol{x}\right)}{\nabla_\perp^2}~,
\ee
with nonvanishing field strength only at $x^-=0$. 
Likewise, the current of the incoming shockwave with the transverse color charge distribution $\rho_L(\bsx)$ moving in the negative $z$ direction:
\be
J_{\mu} = g\delta_{\mu+} \delta(x^+)\rho_L\(\bsx\)~.
\ee
For $t>0$, in the dilute-dilute approximation, one simply linearizes the YM equations to linear order in the sources $\rho_H$ and $\rho_L$ and solves for the radiation field $a_\mu$. In light cone gauge $a_+=0$, the physical components of the gauge field can be expressed as  
\begin{align}
    &\square a_{i,c} = -g^3\(\Theta(x^+)\Theta(x^-) \p_i\(\frac{\rho_H}{\nabla_\perp^2}\rho_L\)-2\delta(x^+)\delta(x^-) \frac{\rho_H}{\nabla_\perp^2} \frac{\p_i\rho_L}{\nabla_\perp^2}\)T^a T^b f_{abc}~,
\end{align}
whose Fourier transform  is 
\begin{align}
    \label{eq:gauge-Lipatov}
    a_{i,c}(k) &= -\frac{2ig^3}{k^2+i\epsilon k^-}\int \frac{d^2\bsq_2}{(2\pi)^2}\(q_{2i}-k_i\frac{\bsq_2^2}{\bsk^2}\)\frac{\rho_H}{\bsq_1^2}\frac{\rho_L}{\bsq_2^2}T^a T^b f_{abc}~.
\end{align}
In this expression,  $1/k^2$ corresponds to the propagator of the emitted gluon,  $1/\bsq_1^2$ and $1/\bsq_2^2$ are the propagators of  reggeized gluon propagators 
emitted by the classical sources, and the Lipatov vertex (in light cone gauge) is the term in the parenthesis. (This expression, {\it a la} LSZ, is related to the inclusive amplitude by amputating the external gluon propagator and placing it on-shell.)

In \cite{Blaizot:2004wu,Gelis:2005pt}, it was shown that the Lipatov vertex is also contained in the classical YM solutions in the dilute-dense scattering regime: 
\begin{align}
\label{eq:dilute-denseQCD}
a_i(k)  =  -\frac{2ig}{k^2+i\epsilon k^-}\int \frac{d^2\bsq_{2}}{(2\pi)^2} \(q_{2i}-k_i\frac{\bsq_2^2}{\bsk^2}\) \frac{\rho_L(\bsq_{2})}{\bsq_{2}^2}\bigg(U(\bsk+\bsq_{2})-(2\pi)^2 \delta^2(\bsk+\bsq_{2})\bigg)\,.
\end{align}
Here $U(\bsk)$  is the Fourier transform of 
\begin{align}
    U(x^-, \bsx)  \delta\left(x^{+}\right)  = \exp\(ig \int_{-\infty}^{x^-} dz^- \bar{A}_-(z^-, \bsx) \cdot T \)~,
\end{align}
where $\bar{A}_-(z^-, \bsx) $ defined in Eq.~\eqref{eq:gluon-ShockBgnd}. The Wilson line encodes the coherent multiple scattering of the emitted gluon off the dense source $\rho_H$ in Fig. \ref{dilute-dense}; expanding to lowest order in $\rho_H$  recovers the dilute-dilute result in Eq.~\eqref{eq:gauge-Lipatov}. We will henceforth for simplicity focus only on the dilute-dilute case. 
 
The average number of gluons produced in a dilute-dilute shockwave collision is~\cite{Blaizot:2004wu,Gelis:2006yv}  
\begin{eqnarray}
\label{eq:single-gluon-inclusive}
    {\bar n}_g = \int \frac{d^3 p}{(2\pi)^3 2 E_p}\langle | {\cal M}^c (p)|^2\rangle_{\rm \rho_L,\rho_H}\,.
\end{eqnarray}
As noted earlier, the amplitude ${\cal M}^c(p) = p^2 a_i^c(p)$. The expectation value here, for a generic operator  ${\cal O}$ is 
\begin{eqnarray}
\label{eq:dilute-dilute-mean}
    \langle {\cal O} \rangle_{\rho_L,\rho_H} = \int [d\rho_L][d\rho_H]\, W_{\rm Y_L + Y}[\rho_L]\, W_{\rm Y_H-Y}[\rho_H]\,{\cal O}[\rho_L,\rho_H]\,,
\end{eqnarray}
where the l.h.s is computed at rapidity $Y$; this expression can be proven to a so-called leading logarithmic accuracy~\cite{Gelis:2008rw}. Specifically, quantum corrections accompanied by large logarithms $\ln(P^\pm/\Lambda^\pm)$ (where $P^\pm$ denotes the momenta of the colliding shockwaves, and $\Lambda^\pm$, the longitudinal momentum modes specifying the rapidity $Y$), can be absorbed\footnote{In other words, ``slow" gluons that are ``bremsstrahlunged" off the shockwaves (but are nevertheless faster than gluons at the scale of interest) can successively be absorbed into the CGC weight functionals $W$ for each colliding shockwave, as it is boosted to higher energies. To this leading logarithmic accuracy, the shockwaves satisfy the JIMWLK RG equation~\cite{Jalilian-Marian:1997ubg,Iancu:2000hn}.} into the CGC weight functionals $W$.  The computation of an operator ${\cal O}[\rho_L,\rho_H]$ in the argument of the path integral is obtained from solutions of the classical Yang-Mills equations, as a functional of the sources (static on time scales of interest) that contain all the fast bremsstrahlunged charges; these correspond to classical color charge representations.

\begin{figure}[ht]
 \centering    
\includegraphics[scale=1]{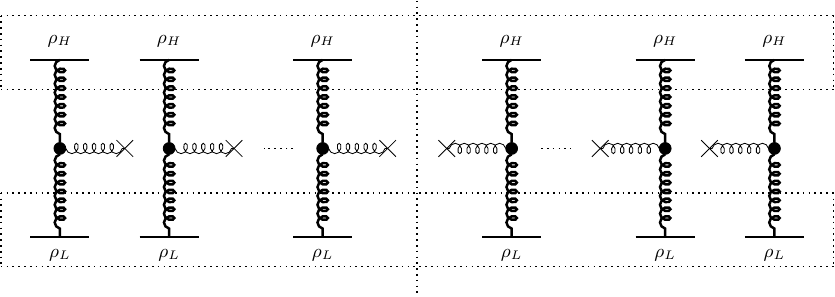}
\caption{Illustration of Eq.~\eqref{eq:kth-moments-gluons} for the n-gluon inclusive multiplicity in the glasma, computed in the dilute-dilute limit of the CGC EFT. The dashed rectangles indicate stochastic averaging over the color sources with the CGC weight functionals $W$. }
\label{fig:glitter}
\end{figure}

Remarkably, Eq.~\eqref{eq:dilute-dilute-mean} also holds\footnote{For further discussion of this equation, and its regime of validity, we refer the reader to 
\cite{Raj:2025hse}. 
} for multiple-gluon emission \cite{Gelis:2008ad}, with the produced matter forming the  nonequilibrium Glasma. The $k$-th factorial moment of the multiplicity distribution can formally be defined as
\begin{eqnarray}
\label{eq:kth-moments-gluons}
    \langle n (n-1)\cdots (n-k-1)\rangle = \int \frac{d^3 p_1}{(2\pi)^3 2 E_{p_1}}\cdots \frac{d^3 p_k}{(2\pi)^3 2 E_{p_k}} \left \langle \frac{d^k N}{d^2 p_{\perp,1}dy_1\cdots d^2 p_{\perp,k}dy_k}\right \rangle_{\rho_L,\rho_H}\,,
\end{eqnarray}
where $\langle\cdots\rangle$ corresponds to Eq.~\eqref{eq:dilute-dilute-mean}. The objects inside the expectation value are however just $k$-products of the l.h.s of 
Eq.~\eqref{eq:single-gluon-inclusive}. In other words, they are products of $k$ independent $2\rightarrow 3$ gluon emission vertices (as illustrated in Fig.~\ref{fig:glitter}) in the amplitude and in the complex conjugate amplitude, for a given distribution of $\rho_{L,H}$, that are sewn together by averaging over all possible distributions with the weight functionals $W$. This $t$-channel fractionation mechanism for multi-particle production, while highly sub-leading to  $s$-channel (BFKL) fractionation at  low gluon occupancies, becomes the dominant mechanism when occupancies approach $1/\alpha_S$. Nevertheless, the building blocks are  Lipatov vertices and $t$-channel reggeized propagators that can be mapped~\cite{Hentschinski:2018rrf} to the classical fields generated by the sources. 

For a Gaussian weight functional, the combinatorics of the color charge densities in Eq.~\eqref{eq:dilute-dilute-mean} can be worked out explicitly~\cite{Dumitru:2008wn,Dusling:2009ar,Gelis:2009wh};  the n-particle probability distribution is the negative binomial distribution (NBD)~\cite{DeWolf:1995nyp,Dremin:2000ep}
\begin{eqnarray}
\label{eq:NBD-QCD}
    P_{n;r} = \frac{\Gamma(n+r)}{\Gamma(r)\Gamma(n+1)}\frac{{\bar n}^n r^r}{({\bar n}+r)^{{ n}+r}}\,,
    \label{eq:NBD-1}
\end{eqnarray}
where $\bar{n}$ is the mean of the distribution and $r$ is defined as~\cite{Gelis:2009wh}
\begin{eqnarray}
\label{eq:NBD-parameter}
    r = \zeta \frac{(N_c^2-1) S_\perp Q_S^2}{2\pi}\,.
\end{eqnarray}
Here $\zeta$ is a nonperturbative $O(1)$ constant and $S_\perp$ is the transverse overlap area of the shockwaves at a fixed impact parameter. The  NBD distribution was employed previously to fit multiplicity distributions with $r$ as a phenomenological parameter~\cite{Dremin:2000ep}; here, it is computed in the CGC EFT dilute-dilute approximation. The variance of the distribution, for instance can be expressed as 
\be
\sigma^2 = {\bar n} + \frac{{\bar n}^2}{r}\,,
\ee
which gives the well-known expression for the variance of a Poisson distribution for $r\rightarrow \infty$, and the significantly broader Bose-Einstein variance for $r=1$. 
The appearance of the saturation scale $Q_S$ in  $r$ in Eq.~\eqref{eq:NBD-parameter} reflects the appearance of nonlinearities that soften the single inclusive 
distribution at transverse momenta $p_T < Q_S$. Though analytical results are only available in the dilute-dilute and dilute-dense cases, numerical simulations of shockwave scattering~\cite{Krasnitz:1998ns} show that the NBD structure persists in this fully nonperturbative case~\cite{Lappi:2009xa,Schenke:2013dpa}.

\section{The Glasma as a squeezed coherent state }

We will demonstrate here that the NBD distribution corresponding to $n$-gluon emission in shockwave scattering can be expressed as a squeezed coherent state. 
We begin by rewriting Eq.~\eqref{eq:NBD-1} as 
\begin{align}
\label{eq:NBD-2}
    P_{n;r} = p^n (1-p)^r \, \frac{\Gamma(n+r)}{\Gamma(r)\Gamma(n+1)}\,,
\end{align}
where $p={\bar n}/({\bar n}+r)$. 
We observe that this distribution has the following limiting properties. For 
\be
\label{poisson-limit}
  r\to\infty,\,\,\,
  p\to0,\,\,\,
  {\rm such\,\, that}\,\,\,
  r\,p=\bar n=\text{constant},
\ee
the distribution~\eqref{eq:NBD-2} becomes the Poisson distribution
\be
\label{eq:poisson}
  P_{n;r} \;\longrightarrow\; P_{n,{\rm Poisson}} = 
  e^{-\bar n}\frac{\bar n^{n}}{n!}\,.
\ee
On the other hand, for $r=1$, Eq.~\eqref{eq:NBD-2} is the  Bose distribution\footnote{In textbooks, the Bose distribution (say, for the grand canonical ensemble) is written as 
\be
P(n) = \(1-e^{-\beta(\eps-\mu)}\)e^{-n\beta(\eps-\mu)
}~,\qquad ~ n = 0, 1, 2, \cdots
\ee
where $\epsilon$ is the energy level and $\mu$ is chemical potential. If we write $e^{-\beta(\eps-\mu)} = \bar{n}$,  the above distribution takes the form of Eq.~\eqref{eq:bose}. Note that, more generally, this is the usual geometric distribution.
}
\begin{align}
\label{eq:bose}
    P_n(r,p) \;\longrightarrow\; P_{n,Bose}(\bar{n}) & = 
  (1-p)p^n \,.
\end{align}

With this in mind, we can directly use Eq.~\eqref{eq:NBD-2} to construct the corresponding state $|z;r\>$ as an expansion in terms of the n-gluon Fock basis with coefficients fine tuned as follows. 
Let $\{|n\rangle\}_{n=0}^{\infty}$ be the bosonic Fock basis.
For any complex $z$ with $|z|<1$ (needed for convergence), we define
\begin{equation}
\label{eq:NBstate}
  |z;r\rangle
  =(1-|z|^2)^{r/2}\!
  \sum_{n=0}^{\infty}
  \sqrt{\binom{n+r-1}{n}}\,
  z^n\,|n\rangle .
\end{equation}
The probability of finding $n$ quanta in this state is\footnote{This state is correctly normalized because of the binomial identity  $\sum_{n\ge0}\binom{n+r-1}{n}|z|^{2n}=(1-|z|^2)^{-r}$, for $|z|<1$.}
\begin{equation}
  P_n = |\langle n|z;r\rangle|^2
       = (1-|z|^2)^{r}\,
         \binom{n+r-1}{n}\,|z|^{2n}.
  \label{eq:NBdistribution}
\end{equation}
which is identical to Eq.~\eqref{eq:NBD-2} for $p=|z|^2$.
The family of states represented by Eq.~\eqref{eq:NBstate}, as noted, continuously interpolates between Bose statistics (for $r=1$) and Poisson statistics
(for $r\to\infty$).

The state $|z;r\>$, which describes the Glasma, can be written as 
\begin{align}
\label{eq: operator-form}
    |z;r\> = (1-|z|^2)^{r/2}e^{z a^\dagger \sqrt{\hat{N}+r}}|0\>~,\qquad |z|<1~.
\end{align}
The proof follows from induction and is given in Appendix A. We can show further that this state is the eigenstate of the 
\emph{deformed annihilation operator} coherent state (see also \cite{Fu:1996ra})
\begin{equation}
  \hat{A} = a\sqrt{\frac{1}{\hat N+r-1}}
  \,\,\,\,,\qquad \hat{N} = a^\dagger a~\,,
  \label{eq:deformed}
\end{equation}
which satisfies the identity
\begin{equation}
\label{eq:deformed-coherent-state}
\hat{A}|z;r\rangle = z\,|z;r\rangle~.
\end{equation}

An interesting question relevant to our earlier discussion is whether Eq.~\eqref{eq: operator-form} is a squeezed coherent state. (For the convenience of readers, we have summarized some relevant properties of coherent and squeezed states in Appendix B.) In the limit of $r\rightarrow \infty$, $\hat{A}\rightarrow a/\sqrt{r}$; therefore, 
$|z; r\rangle$ reduces to a coherent state in this limit. This is of course the  well-known result from quantum optics that coherent states satisfy a Poisson distribution\footnote{The single-mode squeezed state 
$|\eta\> = e^{-\frac12\left(\eta a^2 - \eta^* a^{\dagger 2}\right)} |0\rangle$ gives rise to a negative binomial distribution with parameters $r=1/2$, and $p=\tanh^2|\eta|$. However, $|\eta\>$ does not belong to the NBD family of states; as noted in Appendix B.2, it is a superposition of states of only even particle number while the NBD state contains both even and odd number states.}. 

For $r=1$, $\hat A$ is the Susskind-Glogower (SG) phase operator~\cite{Susskind:1964zz}
\begin{equation}
{\hat A}_{r=1}= \sum_{n=0}^{\infty}|n\rangle\langle n+1|\,,
\end{equation}
which satisfies
\begin{equation}
\label{eq:phase-state}
{\hat A}_{r=1}|\phi\rangle=e^{i\phi}|\phi\rangle\,\,\,{\rm with}\,\,\,|\phi\rangle=\frac{1}{\sqrt{2\pi}}\sum_{n=0}^{\infty}e^{in\phi}|n\rangle \,.
\end{equation}
The Glasma state in Eq.~\eqref{eq: operator-form} for $r\neq 1$ can therefore be understood as a generalized Susskind-Glogower (gSG) state. We note that the Susskind-Glogower phase operator is not unitary. However, as pointed out by Barnett and Pegg~\cite{Pegg:1988wwa}, a small modification to the phase state definition in Eq.~\eqref{eq:phase-state} to restrict the sum over number states to a finite value is sufficient to define a unitary phase operator, which is important for the consistent quantization of such states~\cite{Dirac:1927dy,Susskind:1964zz}.\footnote{We thank Stephen Barnett for an illuminating discussion of his seminal work on this topic.}

A characteristic feature of both coherent states and squeezed states (see Appendix B) is that they saturate the minimum uncertainty relation
\begin{align}
\Delta X^2\,\Delta P^2=\frac{\hbar^2}{4}~,
\end{align}
where $\Delta X^2 = \<X^2\>-\<X\>^2$ and $\Delta P^2 = \<P^2\>-\<P\>^2$, correspond to the variance in position and momentum respectively. (We will henceforth set $\hbar=1$, unless required.) Squeezed states differ from coherent states 
in that the appropriately normalized coherent state values for $\Delta X$ to be  squeezed (or expanded), and conversely for $\Delta P$, such that the product is equal to $1/2$. In other words, if we consider the minimal uncertainty distribution of a coherent state to define a circle in phase space, a squeezed state is an ellipse, where the aspect ratio quantifies the amount of the squeezing. 

What is the corresponding relation satisfied by the gSG state, and how is it squeezed? This is worked out in detail in Appendix C, with the result 
\begin{align}
\label{varx}
\Delta X^2
=
\frac{1}{2}
+
\frac{r p}{1-p}
+
\frac{(1-p)^r}{2}(z^2+z^{*2}) A_2(p)
-
\frac{(1-p)^{2r}}{2}(z+z^*)^2 A_1(p)^2\,.
\\[10pt]
\Delta P^2
=
\frac{1}{2}
+
\frac{r p}{1-p}
-
\frac{(1-p)^r}{2}(z^2+z^{*2}) A_2(p)
+
\frac{(1-p)^{2r}}{2}(z-z^*)^2 A_1(p)^2\,.
\label{varp}
\end{align}
where $A_1$ and $A_2$ are defined, respectively, in Eq.~\eqref{A1-sum} and Eq.~\eqref{A2-sum}, which we simplify and rewrite here as
\begin{align}
\label{A1-real-sum}
A_1(p) &= \sum _{m=0}^{\infty } \sqrt{(m+r)} \binom{m+r-1}{m} p^m \,,\\
A_2(p) &= \sum _{m=0}^{\infty } \sqrt{(m+r)(m+r+1)} \binom{m+r-1}{m} p^{m}\,.
\label{A2-real-sum}
\end{align}
These expressions cannot be written in terms of known special functions. However, the sums converge for $p<1$ for a fixed $r$ and can be evaluated numerically. If we restrict $z$ for now to be real\footnote{We will return to this point shortly.}
\begin{equation}
\label{eq:zero-phase}
z^{2}+z^{*2} = 2p,\qquad
(z+z^{*})^{2} = 4p,\qquad
(z-z^{*})^{2} = 0\,,
\end{equation}
the variances simplify to
\begin{align}
(\Delta X)^{2} &= \frac{1}{2}+\frac{rp}{1-p}
+(1-p)^{r}p\,A_{2}(p)
-2(1-p)^{2r}p\,A_{1}(p)^{2},
\label{eq:dx2_real}\\[8pt]
(\Delta P)^{2} &= \frac{1}{2}+\frac{rp}{1-p}
-(1-p)^{r}p\,A_{2}(p).
\label{eq:dp2_real}
\end{align}
For real $z$, we observe that 
\begin{enumerate}
\item $(\Delta P)^2$ contains no $A_1$ contribution -- it is fully determined by $A_2$ alone.
\item $(\Delta P)^2 < (\Delta X)^2$ for all $r>0$, $p\in(0,1)$, so the
      $\Delta P$-quadrature is always squeezed for real $z$~\cite{Fu:1996ra}. 
\end{enumerate}

Even though $(\Delta P)^{2}$ is squeezed, the gSG state is not a single mode squeezed state since $\Delta X \Delta P > 1/2$. In the limit 
$r\rightarrow \infty$, we do recover a coherent state. To see this, we 
rescale $z$ (recall $p$ was $|z|^2$) as
\be
z = \frac{\alpha}{\sqrt{r}}, \qquad p = \frac{|\alpha|^2}{r}~, \qquad (\alpha~\text{fixed})~.
\ee
The asymptotics are worked out in Appendix C. For real $z$, one finds that\footnote{Note that in the limit $r\rightarrow \infty$, $|\alpha|^2=pr 
\equiv \frac{\bar n r}{\bar n + r}\rightarrow {\bar n}$.}
\begin{align}
    (\Delta X)^2(\Delta P)^2 = \frac{1}{4} + \frac{|\alpha|^6}{32r^4} + O(r^{-6})~,
\end{align}
showing that the leading deviation from the minimal uncertainty state appears at order $1/r^4$ for real $z$ (and order $1/r^2$ for complex $z$ as shown in the Appendix C). 

From Eq.~\eqref{eq:NBD-parameter}, for the Glasma created in central (zero impact parameter) collisions, $S_\perp \propto A^{2/3}$, where $A$ is the nuclear atomic number. Further $Q_S^2\propto A^{1/3}$~\cite{Kowalski:2007rw}, giving $r\propto A\gg 1$ this suggests that for collisions of large nuclei, the Glasma gSG state produced is, to a very good approximation, a squeezed coherent state (since $\Delta P < 1/2$ for $r> 1$). 
\vspace{0.1cm}

\noindent{\it Is the squeezed Glasma state a superfluid?}
\vspace{0.1cm}

In Eq.~\eqref{eq:zero-phase}, we set the phase corresponding to $z$ to zero, to arrive at Eqs.~\eqref{eq:dx2_real} and \eqref{eq:dp2_real}. What is the physical meaning of this? In general, fixing this phase to a definite value, breaks a $U(1)$ symmetry corresponding to number conservation. Fundamentally, this is because the phase operator and the number operator satisfy an uncertainty relation~\cite{Dirac:1927dy,Susskind:1964zz,Pegg:1988wwa,Barnett:01011989}; setting this phase to a fixed value therefore violates the $U(1)$ symmetry. This is typical of superfluids, where the ground state is a condensate with indefinite particle number~\cite{svistunov2015superfluid}. In the context of an {\it overoccupied} $2+1$-D self-interacting scalar field theory, one can show explicitly that its dynamical evolution leads to a number cascade towards the IR, forming a superfluid Bose condensate~\cite{Deng:2018xsk}. The temporal evolution of such overoccupied scalar fields has been shown in numerical simulations to be identical that of the Glasma, strongly suggestive that their dynamics belongs to the same universality class of far-from-equilibrium states~\cite{Berges:2014bba}.

In the CGC/Glasma, it is unclear whether it is meaningful to talk about particle/energy number states of gluons, firstly, because the fields are strongly nonlinear, and secondly, due to the difficulty in defining a gauge invariant order parameter for the condensate. One way to think about the problem is as follows. In the $s$-channel to $t$-channel fractionation (perturbative $\rightarrow$ nonperturbative) transition we outlined in Section 2, the symmetry of large gauge transformations is broken~\cite{Strominger:2017zoo} by the shockwave created in the process, creating different pure gauge vacua on either side of the shockwave~\cite{McLerran:1994vd,Balitsky:1995ub}. The reggeized gluons (solutions of the shockwave classical YM equations), from these symmetry considerations, can be understood as Goldstones of the broken ``BMS" generators of these large gauge tranformations; specifically, these are the $N_c-1$ generators of the $U(1)\times U(1)$ Cartan sub-algebra~\cite{Nair}. The IR dynamics of the shockwave (and of colliding shockwaves) can thus be formulated in this Cartan basis, where a Fock basis representation in terms of non/weakly-interacting Goldstones is realized.\footnote{This novel interpretation may explain why gluons in the shockwave basis must be treated as  non-interacting modes in order to reconcile computations of deeply inelastic scattering final states in different frames~\cite{Mueller:2026fee}.} In this picture, the n-gluon gSG state can be understood as the scattering of reggeized gluons in shockwave scattering to produce a physical gluon, with their interaction mediated by the Lipatov vertex~\cite{Hentschinski:2018rrf}. 

Another possible interesting consequence of the gSG squeezed state description of the Glasma as a superfluid, is the rapid decoherence and thermalization of such states~\cite{Longhi:2024pko}. This would be consistent with experimental observations suggesting rapid thermalization of the QGP. Much work remains to further flesh out these ideas and their practical consequences. 

\section{gSG squeezed state description of multi-graviton radiation in strong gravitational fields}

In Einstein gravity (GR), the scattering of shockwaves can be described by generalizing the Aichelberg-Sexl metric~\cite{Aichelburg:1970dh} to take the form~\cite{Raj:2023irr} 
\be
\label{eq:denseBG}
ds^2 = 2dx^+dx^- -\delta_{ij}dx^i dx^j + f(x^-,\bsx)\(dx^-\)^2 ~,
\ee
with  
\begin{align}
\label{eq:BG-metric}
f(x^-,\bsx) &= 2\kappa^2\mu_H \delta(x^-)\frac{\rho_H(\bsx)}{\square_\perp} = \frac{\kappa^2}{\pi}\mu_H \delta(x^-) \int d^2\bsy ~\ln\Lambda |\bsx-\bsy| \rho_H(\bsy)~,
\end{align}
with $\kappa^2=8\pi G$, and where in analogy to the QCD case, $\rho_H$ is the mass density of the boosted heavy compact gravitational source. The two incoming gravitational shockwaves, along the $z$ axis separated by impact parameter $b$ in the transverse plane with the collision point at $z=t=0$, are sourced by the energy-momentum tensor (for $t<0$),
\begin{align}
\begin{split}
\label{EMtensor2}
T_{\m\n} = &~ \delta_{\mu-}\delta_{\nu-} \m_H \,\delta(x^-) \rho_H(\bsx) +\delta_{\mu+}\delta_{\nu+} \m_L \delta(x^+) \rho_L(\bsx)~,
\end{split}
\end{align}
where $\rho_L$ is the mass density of a light boosted compact gravitational probe. 
Treating the spacetime created by the heavy shockwave as background, small perturbations $h_{\m\n}$ around it can be expressed as 
\begin{align}
\label{perturb}
g_{\m\n} = \bar{g}_{\m\n} + h_{\m\n}~,
\end{align}
where $\bar{g}_{\m\n}$ is the background metric tensor in Eq.~\eqref{eq:BG-metric}. 
Expanding systematically out to order $O(\rho_H\rho_L)$ in light cone gauge $h_{\mu+}=0$, the solutions of Einstein's equations result in the wave equation~\cite{Raj:2023irr} 
\begin{align}
\label{eq:GR-wave-eqn}
&\bar{g}_{--}\p_+^2 \tilde h_{ij}- \square \tilde h_{ij}= \kappa^2\bigg[\(2\p_i\p_j-\square_\perp \delta_{ij}\)\frac{1}{\p_+^2}T_{++}  +2T_{ij}-\delta_{ij}T-\frac{2}{\p_+} \(\p_iT_{+j}+\p_jT_{+i}-\delta_{ij}\p_k T_{+k}\) \bigg]~.
\end{align}
Here $\tilde{h}_{ij} \equiv h_{ij}-\frac12 \delta_{ij} h$ where $h=\delta_{ij}h_{ij}$, $\square$ denotes the d'Alembertian operator, and $T\equiv \delta_{ij}T_{ij}$. One can similarly derive expressions for other components of the gravitational wave tensor generated in the collision. 

Just as for shockwave scattering in QCD, the equations have to be  supplemented with the covariant conservation equations for the current. However, unlike the former,  conservation laws at leading eikonal order are not sufficient to determine the evolution of the energy-momentum tensor. One needs to consider in addition the geodesic motion of the ultrarelativistic distribution of particles comprising the light object (with mass density $\rho_L$) in the shockwave background of the heavy shockwave. With this further input, one can solve Eq.~\eqref{eq:GR-wave-eqn} to obtain the result~\cite{Raj:2023iqn,Raj:2025hse} 
\begin{align}
\label{eq:hijfinal}
    \tilde{h}_{ij}^{(2)}(k) = \frac{2\kappa^3\mu_H\mu_L}{k^2+i\epsilon k^-} \int \frac{d^2\bsq_2}{\(2\pi\)^2}\, \Gamma_{ij}(\bsq_1,\bsq_2) \frac{\rho_H}{\bsq_1^2}\frac{\rho_L}{\bsq_2^2} ~.
\end{align}
The structure of this result can be compared to Eq.~\eqref{eq:gauge-Lipatov}, where now $\Gamma_{ij}$ is the gravitational Lipatov vertex,
\begin{align}
\label{eq:GR-Lipatov}
\Gamma_{ij}(\bsq_1,\bsq_2)&=2\(q_{2i}-k_i\frac{\bsq_{2 }^{2}}{\bsk^{2}}\) \(q_{2j}-k_j\frac{\bsq_{2 }^{2}}{\bsk^{2}}\) - 2k_ik_j\frac{\bsq_{1}^2\bsq_{2}^2}{\bsk^4}\,.
\end{align}
Indeed, by replacing $g\rightarrow \kappa$, color charge densities with mass densities $\rho_{H,L}\rightarrow \mu_{H,L}\,\rho_{H,L}$, and the QCD Lipatov vertex $C_i\rightarrow \Gamma_{ij}$ in Eq.~\eqref{eq:gauge-Lipatov}, one recovers the above result as the double copy of the QCD one. This double copy of the  radiative vertices can be  expressed in the  general covariant form,
\begin{align}
    \Gamma_{\mu\nu}(\bsq_1,\bsq_2) = \frac{1}{2}\left(C_\mu C_\nu - N_\mu N_\nu\right)\,,
\end{align}
showing that the gravitational Lipatov vertex is proportional to a bilinear of its QCD counterpart, and that of  $N^\mu$, which is the QED bremsstrahlung vertex multiplied by a scalar function~\cite{Lipatov:1982it,Raj:2025hse}. Note that both Lipatov vertices are gauge covariant; their scalar product is gauge invariant. Further, as in the QCD case, $1/\bsq_1^2$ and $1/\bsq_2^2$ in Eq.~\eqref{eq:hijfinal} are reggeized graviton propagators emitted by the classical sources. 

These results in the CGC inspired shockwave formalism are an alternative formulation\footnote{For other recent related work in this context, see \cite{Rothstein:2024nlq,Alessio:2025isu,Alessio:2026bdi}.} of the 2-D reggeon high energy EFT formalism for QCD and gravity introduced by Lipatov~\cite{Lipatov:1982it,Lipatov:1991nf}, and subsequently developed into a powerful framework for trans-Planckian gravitational scattering by Amati, Ciafaloni, Veneziano~\cite{Amati:1987uf,Amati:1992zb,Amati:1993tb,Amati:2007ak}, and collaborators-for a nice recent review, see \cite{DiVecchia:2023frv}. The building blocks in this formalism, as in ours, are the Lipatov vertices and reggeized propagators. 

It is therefore reasonable to assume that gravitational radiation from $t$-channel fractionation proceeds in a similar way in the GR case~\cite{Ciafaloni:2015xsr,Ciafaloni:2017ort}, as in QCD. In analogy to QCD, from a QFT perspective, the scale corresponding to the distribution of sources is the Schwarzchild radius $R_S$. An important question is what the 
NBD parameter $r$ is. As in the QCD case discussed in \cite{Gelis:2009wh}, it will be determined by the IR/UV behavior of Eq.~\eqref{eq:hijfinal}. It is reasonable to therefore assume $r\propto R_s/b$, where $R_S$  and $b$ (the impact parameter at which the laser-like $n$-graviton radiation occurs) are the corresponding UV and IR regulators. As we noted previously, $b\gtrapprox R_S$ in the Lipatov radiation regime, so one anticipates $r\sim 1$, as in a Bose distribution. However the prefactor ($\zeta$ in Eq.~\eqref{eq:NBD-parameter}) is unknown; further, the corrections in the dilute-dense regime (under investigation in \cite{FRV}) can be significant. 

Therefore, to estimate the magnitude of squeezing in the gSG gravitational radiation state, we will treat $r$ as a free parameter and explore the possible consequences. To extract the squeezing parameter, we parametrize the variances as
\begin{equation}
\label{eq:param}
(\Delta X)^{2} = \left(\delta+\tfrac{1}{2}\right)e^{-2\xi}\,,
\qquad
(\Delta P)^{2} = \left(\delta+\tfrac{1}{2}\right)e^{+2\xi}\,,
\end{equation}
which gives
\begin{equation}
\label{eq:squeeze-delta}
    \sqrt{(\Delta X)^{2}(\Delta P)^{2}} = \frac12 + \delta\,.
\end{equation}
We can thus write 
\begin{align}
\xi = \frac{1}{4}\ln\frac{(\Delta P)^{2}}{(\Delta X)^{2}}\,,
\label{eq:squeeze_param}
\end{align}
and substituting Eqs.~\eqref{eq:dx2_real} and \eqref{eq:dp2_real} into
Eq.~\eqref{eq:squeeze_param}, results in  
\begin{equation}
\xi = \frac{1}{4}\ln\left(
\frac{\dfrac{1}{2}+\dfrac{rp}{1-p}-(1-p)^{r}p\,A_{2}(p)}
{\dfrac{1}{2}+\dfrac{rp}{1-p}+(1-p)^{r}p\,A_{2}(p)
-2(1-p)^{2r}p\,A_{1}^{2}(p)}\right)\,.
\label{eq:s_full}
\end{equation}
Since $(\Delta P)^{2}<(\Delta X)^{2}$ for real $z$, we have $\xi<0$: the $P$-quadrature is squeezed and $X$-quadrature is anti-squeezed, which, as discussed previously, is what we would want for an observable gravitational signal of quantum noise. We will now analyze the $\delta$, $\xi$ parameter space to see if we can get large squeezing while being close to the minimal uncertainty bound. 

\subsection{The $\delta$, $\xi$ parameter space}
The three conditions we wish to satisfy simultaneously are
\begin{enumerate}
\item[\textbf{(C1)}] $\bar{n} = rp/(1-p)$ is finite,
\item[\textbf{(C2)}] $|\xi|$ is arbitrarily large,
\item[\textbf{(C3)}] $\delta$ is arbitrarily small (near-minimum uncertainty).
\end{enumerate}
First note that the sum of the two variances in Eq.~\eqref{eq:param} is
\begin{equation}
(\Delta X)^{2}+(\Delta P)^{2}
= 2\!\left(\tfrac{1}{2}+\delta\right)\cosh(2\xi).
\label{eq:sum_param}
\end{equation}
From the exact variance formula for the NBD state at real $z$
(see Eqs.~\eqref{eq:dx2_real} and \eqref{eq:dp2_real}), we have 
\begin{equation}
(\Delta X)^{2}+(\Delta P)^{2} = 1+2\bar{n}-Q\,,\,\,{\rm with}
\,\,\, Q \equiv 2(1-p)^{2r}p\,A_{1}(p)^{2}\geq 0.
\label{eq:sum_nbd}
\end{equation}
Equating Eqs.~\eqref{eq:sum_param} and \eqref{eq:sum_nbd}, and using $Q\geq 0$, gives 
\begin{equation}
2\!\left(\tfrac{1}{2}+\delta\right)\cosh(2\xi)
= 1+2\bar{n}-Q
\leq 1+2\bar{n}\,,
\label{eq:master_ineq}
\end{equation}
and therefore, 
\begin{equation}
\label{eq:master}
(1+2\delta)\cosh(2\xi) \leq 1+2\bar{n}~.
\end{equation}
Solving Eq.~\eqref{eq:master} for $\delta$ and $|\xi|$, respectively, then gives
\begin{align}
\delta &\leq \frac{1+2\bar{n}}{2\cosh(2\xi)}-\frac{1}{2}
\equiv \delta_{\max}(\xi,\bar{n})\,,
\label{eq:delta_ub}\\[6pt]
|\xi| &\leq \frac{1}{2}\cosh^{-1}\!\left(\frac{1+2\bar{n}}{1+2\delta}\right)
\equiv \xi_{\max}(\delta,\bar{n})\,.
\label{eq:xi_ub}
\end{align}
Two observations immediately follow.
\begin{enumerate}
\item $\delta_{\max}$ is a {\it decreasing} function of $|\xi|$ at fixed
$\bar{n}$.  As squeezing increases, the upper bound on $\delta$ falls. This shows that conditions \textbf{(C2)} and \textbf{(C3)} are {\it not} in tension with each other -- more squeezing automatically enforces less excess noise.
\item Further, enforcing that $\delta_{\max}\geq 0$, or equivalently,  $\cosh(2\xi)\leq 1+2\bar{n}$, results in an upper bound on $|\xi|$ for finite $\bar{n}$:
\begin{equation}
|\xi| \leq \frac{1}{2}\cosh^{-1}(1+2\bar{n}) < \infty,
\end{equation}
Conditions \textbf{(C1)} and \textbf{(C2)} are therefore \emph{irreconcilable} -- we cannot have arbitrarly high squeezing at finite $\bar{n}$. Conditions \textbf{(C1)} and \textbf{(C2)} cannot hold
simultaneously for any $(r,p)$ in the NBD family.
\end{enumerate}
The maximum squeezing
achievable\footnote{The equality in Eq.~\eqref{eq:master} requires $Q=0$.  However, since $p>0$ and $r>0$,
each term in the sum
$A_{1}(p)$ is strictly positive, so $Q>0$ strictly and the bound in Eq.~\eqref{eq:xi_max}
is never saturated.  The maximum squeezing achievable by a
nontrivial NBD state at fixed $\bar{n}$ and $\delta=0$ is strictly
less than $|\xi|_{\max}$.
} in the NBD state at fixed $\bar{n}$ and $\delta=0$ is
\begin{equation}
|\xi|_{\max} = \frac{1}{2}\cosh^{-1}(1+2\bar{n}),
\label{eq:xi_max}
\end{equation}
which grows logarithmically in $\bar{n}$ for large $\bar{n}$:
$|\xi|_{\max}\approx \frac{1}{2}\ln(4\bar{n})$.
Thus it is possible to achieve small $\delta$ and large squeezing $\xi$ for large occupancies ${\bar n}$.  

We will now explore further the parameters $r$ and $p$ for which one has small $\delta$ and large $\xi$. We find that this corresponds to the regime where $r > 1$ and $p \to 1$. Setting $\epsilon = 1 - p$, the asymptotic forms of the variances are (see Appendix C for details)
\begin{equation}
(\Delta X)^{2} \approx \frac{C(r)}{\epsilon}\,,
\qquad
(\Delta P)^{2} \approx D(r)\,\epsilon\,,
\label{eq:asymp_corrected}
\end{equation}
where
\begin{equation}
C(r) = 2r - \frac{2\,\Gamma\!\left(r+\tfrac{1}{2}\right)^{2}}{\Gamma(r)^{2}}
~\,,
\qquad
D(r) = \frac{1}{2} + \frac{1}{8(r-1)}~\,.
\label{eq:CD}
\end{equation}
As a consequence, 
\begin{align}
|\xi| &= \frac{1}{4}\ln\frac{(\Delta X)^{2}}{(\Delta P)^{2}}
        \approx -\frac{1}{2}\ln\epsilon
        + \frac{1}{4}\ln\frac{C(r)}{D(r)},
\label{eq:xi_asymp}\\[6pt]
\delta &= \sqrt{(\Delta X)^{2}(\Delta P)^{2}} - \frac{1}{2}
        \approx \sqrt{C(r)\,D(r)} - \frac{1}{2}.
\label{eq:delta_asymp}
\end{align}

Eq.~\eqref{eq:xi_asymp} shows that the dominant contribution to
$|\xi|$ is $\frac{1}{2}|\ln\epsilon|$, which depends only on $p$, while
the correction $\frac{1}{4}\ln(C/D)$ is $O(r^{-1})$ for large $r$.  We see from Eq.~\eqref{eq:delta_asymp} that
$\delta$ only depends on $r$.  The two quantities of interest $\xi$ and $\delta$ therefore
depend independently, respectively,  on $p$ and $r$ for  $r> 1$ and $p\to1$. 
In the limit of large $r$ we have that 
\begin{equation}
|\xi|\approx \frac12 |\log(1-p)|~,\qquad \delta \approx \frac{1}{32r}~.
\label{eq:delta_large_r}
\end{equation}
Note that since the prefactor in the expression for $\delta$ is very small, the NBD parameter $r$ does not have to be particularly large for $\delta$ to be small. Equivalently, given a target squeezing $|\xi|_{0}$ and a target $\delta_{0} > 0$, the above equations prescribe the appropriate NBD parameters in the regime of our interest:
\begin{equation}
r = \frac{1}{32\,\delta_{0}}\,,
\qquad
p = 1 - e^{-2|\xi|_{0}}\,.
\label{eq:optimal_params}
\end{equation}
Also, let us look at the mean number $\bar{n}=\frac{rp}{1-p}$ at the above point:

\begin{figure}[ht]
 \centering    
\includegraphics[scale=0.75]{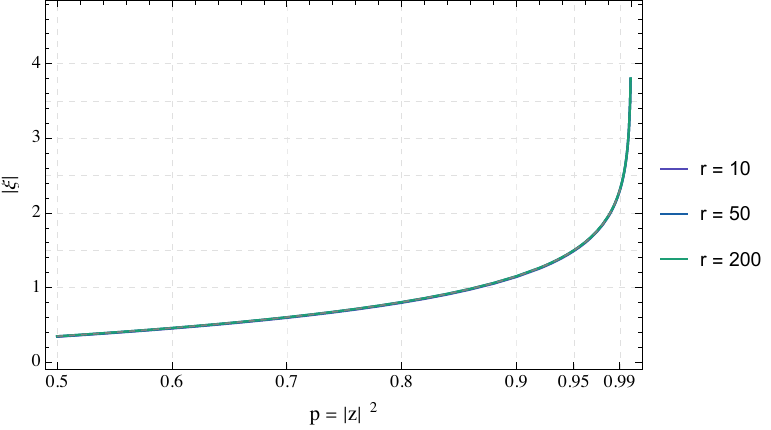}
\caption{The squeezing parameter
$|\xi|$ as a function of $p$, computed by substituting Eqs.~\eqref{eq:dx2_real} and \eqref{eq:dp2_real} in Eq.~\eqref{eq:squeeze_param}. It grows exponentially as $p\to 1$, approaching the analytic asymptote
$\frac{1}{2}|\ln(1-p)|$ for all $r$.}
\label{fig:xip-plot}
\end{figure}

\begin{figure}[ht]
 \centering    
\includegraphics[scale=0.75]{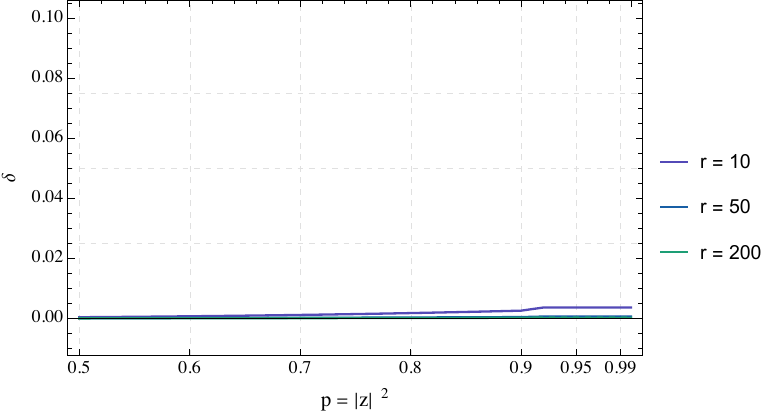}
\caption{Plot of $\delta$ as a function of $p$, for various values of $r$, computed by substituting Eqs.~\eqref{eq:dx2_real} and \eqref{eq:dp2_real} in Eq.~\eqref{eq:squeeze-delta}. As $r$ increases, the value of $\delta$ remains close to zero as $p\to 1$. The seeming jump in the plot for $r=10$ at $p=0.92$ is purely a numerical artifact since we kept just the leading asymptotics to do the matching. Including higher order terms in the asymptotic expansion will lead to a smooth matching.}
\label{fig:deltap-plot}
\end{figure}

\begin{equation}
\bar{n} = \frac{rp}{1-p} \approx \frac{e^{2|\xi|_{0}}}{32\,\delta_{0}} ~.
\end{equation}
This result suggests that the required mean particle number grows \emph{exponentially} in the target squeezing parameter and \emph{inversely} in the
target $\delta_0$. This is not a limitation but rather the expected
physical scaling: the NBD nonlinear coherent state can achieve arbitrary
squeezing at arbitrary proximity to minimum uncertainty, provided the mean particle number is sufficiently large. 

These analytical findings are supplemented by numerically analyzing the variances $(\Delta X)^2$ and $(\Delta P)^2$. The results are shown in Figs.~\ref{fig:xip-plot} and \ref{fig:deltap-plot}. The plots were generated by splitting the $p$-range at a crossover point $p_{\text {cross }}$ that we choose to be $0.92$. Below this crossover, the series for $A_1$ and $A_2$ are evaluated directly with $2500$ terms in the sums. Above the crossover point, the exact asymptotic expressions $(\Delta x)^2 \approx C(r) /(1-p)$ and $(\Delta p)^2 \approx D(r)(1-p)$ are used in place of the series expression, producing smooth curves all the way to $p\to1$. We observe that the two regions join continuously since the values from the asymptotic formulas and those from the numerical estimation of the sums agree to a good degree.

For a typical gravitational wave measured by LIGO, ${\bar n} = 4\cdot 10^{36}$~\cite{Tobar:2023ksi}. The squeezing parameter from our computation is then $|\xi|\sim 42$; this enhances the amplitude for quantum noise from $\sim 10^{-35}$m to $\sim 10^{-17}$ m, which is significantly above the required sensitivity. Since the gravitational flux is proportional to the square of the strain amplitude of the source~\cite{Toccacelo:2026hcz}, Lipatov radiation from primordial sources corresponding to much larger strain amplitudes, such as supermassive black hole mergers, or extreme mass ratio inspirals, can further bring signals of quantum noise into detectable ranges. 

Whether the noise measured is truly quantum is however a much more difficult question. A  nice discussion of this issue can be found in \cite{Carney:2024dsj}. The gSG squeezed state characterizing both the Glasma and 
gravitational radiation in shockwave collisions is  fundamentally a quantum state because it is an eigenstate of the Barnett-Pegg phase operator. However, the corresponding probability distribution is a negative binomial distribution (satisfying super-Poisson statistics), which could be either classical or quantum, and therefore cannot be easily distinguished from a distribution of purely classical origins. In contrast, sub-Poisson statistics are a smoking gun for quantum effects. Despite significant recent developments in novel proposals to detect single gravitons~\cite{Tobar:2023ksi,Shenderov:2024rup,Kharzeev:2025lyu,Palessandro:2024ria}, these too may be inconclusive, as argued in \cite{Carney:2024dsj}. An interesting question in this regard is whether an unambiguous sub-Poissonian signal due to quantum corrections to the gSG state can be detected by measuring 
correlations between ``graviton clicks" in two or more such spatially separated detectors~\cite{Manikandan:2025lfx}, analogously to measurements of Hanbury-Brown--Twiss (HBT) correlations. Baym's Zakopane lectures~\cite{Baym:1997ce}, and a more recent review by Aspect~\cite{aspect2020hanburry} provide  excellent introductions to HBT interferometry from sub-femtometer scales to cosmological ones.

\section{Summary and Outlook}
We discussed here a double copy of gluon and graviton radiation from shockwave collisions in QCD and Einstein gravity, respectively. The gravitational effective Lipatov vertex controlling the emission of gravitons from the interaction of two reggeized gravitons emitted in the shockwave collision is proportional to a bilinear of its QCD counterpart. In QCD, multi-particle production in the CGC EFT can be understood as $t$-channel iterations of the underlying $2\rightarrow 3$ process, with the $n$-particle probability obtained by averaging over the color charge distributions of fast (``large x") partons in each of the shockwaves. This distribution is a negative binomial distribution (NBD), and we showed that it can be described as a generalized Susskind-Glogower (gSG) intermediate phase squeezed coherent state, characterized by the average multiplicity ${\bar n}$ and the NBD parameter $r$. Both of these quantities are computed in the CGC EFT. 

We discussed the possibility that the non-equilibrium Glasma formed as a result of such ``laser-like" copious gluon production (with ${\bar n}\sim Q_S^2 S_\perp/\alpha_S$) is a superfluid state, as is the case for overoccupied ultracold Bose gases. Further, if the state is a strongly squeezed gSG state, one possible consequence  (discussed previously for far-from-equilibrium dynamics in condensed matter physics) is the rapid decoherence and thermalization of Glasma. This possibility is very intriguing in the context of the rapid thermalization and strong flow observed for the QGP formed in ultrarelativistic heavy-ion collisions. 

The shockwave computation in gravity of single inclusive graviton radiation in the Lipatov regime follows exactly analogously to the QCD case. It is therefore very plausible that the resulting n-graviton spectrum follows an NBD distribution as for the Glasma, and is similarly a gSG squeezed state. Since the NBD parameter $r$ in this case is unknown, we explored the behavior of the squeezing parameter $\xi$ for small deviations $\delta$ from the minimum uncertainty state, as a function of ${\bar n}$ and $r$. For $r>1$, and 
${\bar n}\gg r$, large values of $\xi$ are feasible. In particular, we find that $\xi\propto ln({\bar n})$. Since ${\bar n}$ can be very large ($\sim 10^{36}$ for gravitational waves detected by LIGO), our result is very suggestive that variances due to fluctuations on the quantum scale  $L_{\rm Planck}\sim 10^{-35}$ meters can be significantly enhanced towards scales that are accessible at current and future gravitational wave detectors. 

Our conclusion relies on the assumption that the double copy motivated NBD/gSG description holds for multi-graviton radiation in strong gravitational fields. It will therefore be important to extend our shockwave computations of Lipatov radiation in the dilute-dilute and (ideally) dilute-dense analytical frameworks to two-and multi-graviton emissions. In addition to confirming that this correspondence is robust, it will be useful to examine closely how two-graviton correlations can be modified by Hawking-type quantum effects, and whether these lead to sub-Poisson statistics. Not least, since our computations are for ultrarelativistic shockwave collisions, it will be important to understand the $1/\gamma$ corrections to this picture, where $\gamma$ is the Lorentz factor.

\section*{Acknowledgements}
R.V would like to thank Stephen Barnett, Nic Westerberg, and the Quantum Optics theory group at the Univ. of Glasgow for valuable discussions that have influenced this work.

R.V. is supported at Stony Brook by the Simons Foundation as a co-PI under Award number 994318 (Simons Collaboration on Confinement and QCD Strings). H.R. is a Simons Foundation Post-doctoral Fellow at Stony Brook supported under Award number 994318. A.M.S. is  supported by the U.S. Department of Energy grant No. DE-SC-0002145. R.V. is supported by the U.S. Department of Energy, Office of Science under contract DE-SC0012704. A.M.S and R.V. are supported  within the framework of the Saturated Glue (SURGE) Topical Collaboration in Nuclear Theory.  R.V. thanks the UK Royal Society and the Wolfson Foundation for a Visiting Fellowship and the Higgs Center at the Univ. of Edinburgh for their kind hospitality.

\section*{Appendix A: Proofs of Eqs.~\eqref{eq: operator-form} and \eqref{eq:deformed-coherent-state}}
\label{section:AppendixB}

To prove Eq.~\eqref{eq: operator-form}, first note by induction it follows that (proof given below)
\be
\label{op-eq-1}
\(a^\dagger \sqrt{\hat{N}+r}\)^n|0\> = \sqrt{\frac{\Gamma(n+r)}{\Gamma(r)}}\sqrt{n!}|n\>~.
\ee
It then follows that
\begin{align}
    \sum_{n=0}^{\infty}
  \frac{z^n}{n!}\(a^\dagger \sqrt{\hat{N}+r}\)^n \,|n\> = \sum_{n=0}^{\infty}
  \sqrt{\binom{n+r-1}{n}}\,
  z^n\,|n\rangle 
\end{align}
which readily gives the relation in Eq.~\eqref{eq: operator-form}.

\subsubsection*{Proof of \eqref{op-eq-1} by induction}

We first prove the general relation
\be
\label{op-proof-2}
\sqrt{\hat{N}+r}~ |n\>  = \sqrt{n+r} ~|n\>~.
\ee
To do so, denote $\hat{O}\equiv \hat{N}+r$, for which we have 
\be
\label{op-proof-3}
\hat{O}|n\> = (\hat{N}+r) |n\>  = (n+r) |n\>~.
\ee
We want to find the action of an operator $\hat{Q}$ on $|n\>$ such that $\hat{Q}^2 = \hat{O}$. Note that $\hat{O}$ is diagonal on the Fock basis. Its matrix elements takes the following form
\begin{equation}
[\hat{O}] =
\begin{pmatrix}
r & 0 & 0 & 0 & \cdots \\
0 & 1+r & 0 & 0 & \cdots \\
0 & 0 & 2+r & 0 & \cdots \\
0 & 0 & 0 & 3+r & \cdots \\
\vdots & \vdots & \vdots & \vdots & \ddots
\end{pmatrix}~.
\end{equation}
Now because the square root of a diagonal matrix, is simply the square root of the individual diagonal elements, the matrix elements of $\hat{Q}$ take the form
\begin{equation}
[\hat{Q}] =
\begin{pmatrix}
\sqrt{r} & 0 & 0 & 0 & \cdots \\
0 & \sqrt{1+r} & 0 & 0 & \cdots \\
0 & 0 & \sqrt{2+r} & 0 & \cdots \\
0 & 0 & 0 & \sqrt{3+r} & \cdots \\
\vdots & \vdots & \vdots & \vdots & \ddots
\end{pmatrix}~.
\end{equation}
Therefore, $\hat{Q}|n\> = \sqrt{n+r}~|n\>$. This proves Eq.~\eqref{op-proof-2}. 

We will now use this in the proof of Eq.~\eqref{op-eq-1} by induction. Assume that Eq.~\eqref{op-eq-1} is true for an arbitrary but fixed positive integer $m$. We can then show that it is true for $m+1$ as well:
\begin{align}
    &\(a^\dagger\sqrt{\hat{N}+r}\)^{m+1}|0\> =\(a^\dagger\sqrt{\hat{N}+r}\)\(a^\dagger\sqrt{\hat{N}+r}\)^{m}|0\> = \(a^\dagger\sqrt{\hat{N}+r}\)\sqrt{\frac{\Gamma(m+r)}{\Gamma(r)}}\sqrt{m!}|m\>\no\\[5pt]
    &=a^\dagger\sqrt{\frac{\Gamma(m+r)}{\Gamma(r)}}\sqrt{m!}\(\sqrt{\hat{N}+r}\)|m\>=a^\dagger\sqrt{\frac{(m+r)\Gamma(m+r)}{\Gamma(r)}}\sqrt{m!}|m\>~\no\\[5pt]
    &=\sqrt{\frac{(m+r)\Gamma(m+r)}{\Gamma(r)}}\sqrt{m!}\sqrt{m+1}|m+1\>=\sqrt{\frac{\Gamma(m+r+1)}{\Gamma(r)}}\sqrt{(m+1)!}|m+1\>~,
\end{align}
which is the desired result proving Eq.~\eqref{eq: operator-form}.
Note that we used the identity Eq.~\eqref{op-proof-2} for  $\sqrt{\hat{N}+r}|m\>$ in the last term of the second line, and subsequently, $a^\dagger |m\> = \sqrt{m+1} |m+1\>$. 

To prove Eq.~\eqref{eq:deformed-coherent-state}, we begin by using the identities:
\begin{align}
    a|n\> &= \sqrt{n}|n-1\>~,\no\\
    (\hat{N}+r-1)^{-1/2}|n\> &= (n+r-1)^{-1/2}|n\>~,\no\\
    \hat{A} |n\> &= \frac{\sqrt{n}}{\sqrt{n+r-1}} |n-1\>~.
\end{align}
This gives
\begin{align}
A|z;r\rangle
&= (1-|z|^2)^{r/2}
   \sum_{n=1}^{\infty}
   \sqrt{\binom{n+r-1}{n}}\,
   z^n\,
   \frac{\sqrt{n}}{\sqrt{n + r - 1}}\,|n-1\rangle.
\end{align}
Relabeling the summation index $n \mapsto n+1$ yields
\begin{align}
A|z;r\rangle
&= (1-|z|^2)^{r/2}
   \sum_{n=0}^{\infty}
   z^{\,n+1}
   \sqrt{\binom{n+r}{n+1}}\,
   \frac{\sqrt{n+1}}{\sqrt{n + r}}\,|n\rangle.
\end{align}
We now use the binomial identity
\begin{equation}
\binom{n+r}{n+1}\,\frac{n+1}{n+r}
= \binom{n+r-1}{n}.
\end{equation}
Hence, the coefficient of $|n\rangle$ becomes
\begin{equation}
(1-|z|^2)^{r/2}\,z^{\,n+1}
\sqrt{\binom{n+r-1}{n}}|n\>
= z\,(1-|z|^2)^{r/2}\,z^n
\sqrt{\binom{n+r-1}{n}}|n\>.
\end{equation}

But this is simply $z$ times the coefficient of $|n\rangle$ in the original state $|z;r\rangle$.
Therefore we have established Eq.~\eqref{eq:deformed-coherent-state}.

\section*{Appendix B: Coherent states and Squeezed states}

An excellent short pedagogical review of coherent and squeezed states is given in ~\cite{Bagchi:2020hvk}. 
Here we summarize some of the relevant results for this work for the convenience of readers; interested readers should however consult this review and the references within.

\subsection*{B.1: `Linear' Coherent State in a Harmonic Oscillator}
Some basic definitions:
\begin{align}
H &= \frac{\hat{P}^2}{2m} + \frac{1}{2} m\omega^2 \hat X^2\,\,\,;\,\,\,
[\hat X,\hat P] = i\hbar\,\,\,;\,\,\,
\hat X = \sqrt{\frac{\hbar}{2m\omega}}\,(a+a^{\dagger}),\quad \hat P = -i\sqrt{\frac{m\hbar\omega}{2}}\,(a-a^{\dagger}),\\[10pt]
&[a,a^{\dagger}]=1,\quad N=a^{\dagger}a,\quad H=\hbar\omega\left(N+\tfrac12\right).
\end{align}
Fock basis: 
\begin{equation}
|n\rangle = (a^{\dagger})^n/\sqrt{n!}\,|0\rangle\,\,,\,\,  a|n\rangle=\sqrt{n}\,|n-1\rangle\,\,, \,\,a^{\dagger}|n\rangle=\sqrt{n+1}\,|n+1\rangle~\,\,,\,\,\langle n|n\rangle = 1~.
\end{equation}
The last equality follows from $a^n (a^\dagger)^n = (N+1)(N+2) \cdots (N+n)$ with $N$ being the number operator $N=a^\dagger a$ and $\langle0|0\rangle=1$.

Glauber coherent state states $|\alpha\rangle$ are defined as eigenstates of the annihilation operator $a$
\begin{align}
a|\alpha\rangle = \alpha|\alpha\rangle,\quad \alpha\in\mathbb C~,\qquad \langle\alpha|\alpha\rangle = 1~
\end{align}
which satisfy 
\begin{align}
\langle a\rangle_{\alpha}&\equiv \langle\alpha| a|\alpha\rangle =\alpha,\quad \langle N\rangle_{\alpha}\equiv \langle\alpha| N|\alpha\rangle=|\alpha|^2~,
\end{align}
The states saturate the Heisenberg uncertainty bound. To see this saturation, one can compute the variances (uncertainty) in position and momentum: $\<\alpha|(\Delta X)^2|\alpha\> = \<\alpha|X^2|\alpha\>-\<\alpha|X|\alpha\>^2$, $\<\alpha|(\Delta P)^2|\alpha\> = \<\alpha|P^2|\alpha\>-\<\alpha|P|\alpha\>^2$. For instance, 
\begin{align}
    \<\alpha|X|\alpha\> &= \sqrt{\frac{\hbar}{2m\omega}}\ \<\alpha|(a+a^{\dagger})|\alpha\> = \sqrt{\frac{\hbar}{2m\omega}} (\alpha+\alpha^*)~,\\
    \<\alpha|X^2|\alpha\> &= \frac{\hbar}{2m\omega}\ \<\alpha|(a+a^{\dagger})^2|\alpha\> = \frac{\hbar}{2m\omega} \left((\alpha+\alpha^*)^2+1\right)~,\\[5pt]
    \<\alpha|(\Delta X)^2|\alpha\>&= \frac{\hbar}{2m\omega}~.
\end{align}
and likewise for $ \<\alpha|(\Delta P)^2|\alpha\>$, to give 
\begin{align}
\<(\Delta X)^2\>_\alpha\<(\Delta P)^2\>_\alpha=\frac{\hbar^2}{4}~.
\end{align}
 
Expansion in Fock basis:
\begin{align}
\label{expansion-in-fock-basis}
|\alpha\rangle = e^{-\tfrac12|\alpha|^2}\sum_{n=0}^{\infty}\frac{\alpha^n}{\sqrt{n!}}|n\rangle~,
\end{align}
The probability of finding the coherent state in an $n$-particle Fock state
\begin{equation}
    |\langle n|\alpha\rangle|^2= e^{-|\alpha|^2}\frac{|\alpha|^{2n}}{n!} \; ~\qquad(\text{Poisson distribution with mean }|\alpha|^2 ).
\end{equation}
Displacement operator definition and properties:
\begin{align}
D(\alpha)=e^{\alpha a^{\dagger}-\alpha^* a}\,\,;\,\, D^{\dagger}(\alpha)aD(\alpha)=a+\alpha\,\,;\,\,
D(\alpha+\beta)&=e^{\tfrac12(\alpha\beta^*-\alpha^*\beta)}D(\alpha)D(\beta).
\end{align}
The coherent state is defined via the displacement operator as 
$|\alpha\rangle=D(\alpha)|0\rangle$ and satisfies the 
non-orthogonality and overlap relations,
\begin{align}
\langle \alpha|\alpha'\rangle = \exp\!\left(-\tfrac12|\alpha|^2-\tfrac12|\alpha'|^2+\alpha^*\alpha'\right)=e^{-\tfrac12|\alpha-\alpha'|^2}e^{i \Im(\alpha^*\alpha')}, 
\end{align}
and the completeness relation $\frac{1}{\pi}\int d^2\alpha\;|\alpha\rangle\langle\alpha| = \mathbb{I}$. This is derived using Eq.~\eqref{expansion-in-fock-basis} and the identity $\sum_n |n\>\<n|=\mathbb{I}$. 

\subsection*{B.2: Squeezed Coherent states in Harmonic Oscillator}
Define the single mode squeezing operator 
\begin{align}
\label{eq:single-mode-squeeze}
    S_\eta = e^{-\frac12\left(\eta a^2 - \eta^* a^{\dagger 2}\right)}~,
\end{align}
which generates the squeezed state  $|\eta\> = S_\eta |0\rangle$ from the Fock vacuum. This state can also be represented as an expansion in the Fock basis. In order to do so,  we will disentangle  $S_\eta$ into $SU(1,1)$ basis elements. First define operators
\begin{align}
    K_- = \frac12 a^2~,\qquad K_+ = \frac12 a^{\dagger2}~,\qquad K_0 = \frac12\left(a^\dagger a+\frac12\right)~.,
\end{align}
which generate the $su(1,1)$ algebra: $[K_0,K_\pm]=\pm K_\pm~, [K_-,K_+]=2K_0$. Now define $\eta=|\eta|e^{i\phi}$. Then the $su(1,1)$ disentangled form of $S_\eta$, $S_\eta=e^{-\frac12\left(\eta a^{2}-\eta^{*}a^{\dagger2}\right)}=e^{-(\eta K_- - \eta^* K_+)}$ is 
\begin{align}
S_\eta=
\exp\!\big(\gamma\,K_+\big)\;
\exp\!\big(-2\ln(\cosh |\eta|)\,K_0\big)\;
\exp\!\big(-\gamma^*\,K_-\big)~,
\end{align}
with $\gamma = e^{-i\phi}\,\tanh(|\eta|)$.
In terms of $a,a^\dagger$  the squeezing operator is then 
\begin{align}
e^{-\frac12\left(\eta a^2 - \eta^* a^{\dagger 2}\right)}
=\exp\!\big(\tfrac12 e^{-i\phi}\tanh(|\eta|)\,a^{\dagger2}\big)\,
\exp\!\big(-\ln\!\big(\cosh |\eta|\big)\,(a^\dagger a+\tfrac12)\big)\,
\exp\!\big(-\tfrac12 e^{+i\phi}\tanh(|\eta|)\,a^{2}\big)~.
\end{align}
With this representation, we find
\begin{align}
    |\eta\>=S_\eta|0\rangle =\big(\cosh |\eta|\big)^{-1/2} \exp\!\Big(\tfrac12 e^{-i\phi}\tanh(|\eta|)\,a^{\dagger2}\Big)|0\>
    =\big(\cosh |\eta|\big)^{-1/2} \sum_{n=0}^\infty \frac{\gamma^n \sqrt{(2n)!}}{2^n n!}|2n\>\,,
\end{align}
where we used $a^{\dagger n}|0\> = \sqrt{n!} |n\>$ in the last equality. The probability of finding the state $|\eta\>$ in an $2n$ particle Fock state (by construction) is given by
\begin{align}
    p_n = |\<2n|\eta\>|^2 = (\cosh |\eta|)^{-1} \frac{\tanh^{2n}(|\eta|) (2n)!}{2^{2n} (n!)^2}~.
\end{align}
One can verify that this probability distribution (which is no longer Poisson) is properly normalized, $\sum_{n=0}^\infty p_n = 1$. Further, upon inspection we find that this distribution is actually a negative binomial with parameters $r=1/2$ and $p=\tanh^2|\eta|$. This follows simply from the binomial identity
\begin{align}
    \binom{n+\frac{1}{2}-1}{n}=\frac{(2 n)!}{2^{2 n} (n!)^2}~.
\end{align}
We also present results for various expectation values in this state. First of all since only even Fock states contribute, we have $\langle{a}\rangle = 0$. Consequently, $\langle{X}\rangle = \langle{P}\rangle = 0$. The expectation of the number operator is $\langle{N}\rangle = \sinh^{2}|\eta|$. Further, for real $\eta$ a simple calculation gives
\begin{align}
\langle{a}^{2}\rangle
&= \frac{1}{\cosh|\eta|\tanh|\eta|}
\sum_{n=1}^{\infty}\frac{2n}{4^{n}}\binom{2n}{n}\tanh^{2n}|\eta|
= \frac{1}{2}\sinh(2|\eta|).
\end{align}
The last result uses $\sum_{n=1}^{\infty}\frac{2n}{4^{n}}\binom{2n}{n}p^{n}
=p(1-p)^{-3/2}$ with $p=\tanh^{2}|\eta|$.
Inserting into the quadrature variance formula:
\begin{align}
\langle X^2 \rangle
&=
\frac{1}{2}
\left(
\langle a^2 \rangle + \langle a^{\dagger2} \rangle
+ 2\langle N \rangle + 1
\right)~,\qquad \langle P^2 \rangle
=
-\frac{1}{2}
\left(
\langle a^2 \rangle + \langle a^{\dagger2} \rangle
- 2\langle N \rangle - 1
\right)~,
\end{align} 
gives the result:
\begin{align}
(\Delta X)^{2} &= \tfrac{1}{2}e^{2|\eta|}~,\qquad(\Delta P)^{2} = \tfrac{1}{2}e^{-2|\eta|},
\end{align}
Hence, the squeezing parameter $\xi = \frac{1}{4}\ln\((\Delta P)^{2}/(\Delta X)^{2}\)$ is simply $\xi = -|\eta|$ and the state is
\emph{minimum uncertainty}: $(\Delta X)^{2}(\Delta P)^{2} = \frac{1}{4}$.

\section*{Appendix C: Uncertainty relations for the gSG state}

We will compute here the root mean square product  $\Delta X \Delta P$ (in other words, the uncertainty relation) for the generalized Susskind-Glogower state. 
Defining   
\begin{equation}
\label{eq:gSG-coeff}
c_n = (1-|z|^2)^{r/2}
\sqrt{\binom{n+r-1}{n}}\, z^n\,,
\end{equation}
we can write the gSG state in Eq.~\eqref{eq:NBstate} as 
\begin{equation}
|z;r\rangle = \sum_{n=0}^\infty c_n |n\rangle.
\end{equation}
The expectation value of the number operator is (replacing further $|z|^2=p$),
\begin{align}
\label{eq:expect-number}
&\langle N \rangle\equiv {\bar n}
= \sum_{n=0}^\infty n |c_n|^2 
= (1-p)^r \sum_{n=0}^\infty n \binom{n+r-1}{n} p^n 
= (1-p)^r \cdot \frac{r p}{(1-p)^{r+1}} 
= \frac{r p}{1-p}\,,
\end{align}
or equivalently $p= {\bar n}/({\bar n}+r)$.
To compute $ \Delta X$ and $\Delta P$, we will first compute 
the expectation value of the annihilation operator $a$ in the gSG state. Since
\be
a|n\rangle = \sqrt{n}|n-1\rangle,
\ee
we get
\begin{align}
\langle a \rangle
= \sum_{m,n} c_m^* c_n \langle m|a|n\rangle 
= \sum_{n=1}^\infty c_{n-1}^* c_n \sqrt{n}\,,
\end{align}
and substituting Eq.~\eqref{eq:gSG-coeff} in this expression gives 
\begin{align}
\langle a \rangle
&= (1-p)^r z
\sum_{n=1}^\infty
\sqrt{n\binom{n+r-2}{n-1}\binom{n+r-1}{n}}
\, p^{n-1}.
\end{align}
Defining 
\begin{align}
\label{A1-sum}
A_1(p) &= \sum_{n=1}^\infty \sqrt{n}
\sqrt{\binom{n+r-2}{n-1}\binom{n+r-1}{n}}\, p^{n-1}\,, 
\end{align}
we can rewrite the expectation value as 
\begin{align}
\label{eq:expect-a}
\langle a \rangle
&= (1-p)^r z \,A_1(p).
\end{align}
To the best of our knowledge, it is not possible to express $A_1(p)$ in terms of known functions. 

Likewise, the expectation of $a^2$ in the gSG state gives 
\begin{align}
\label{eq:expect-a2}
\langle a^2 \rangle
&= (1-p)^r z^2 A_2(p)\,,
\end{align}
where 
\begin{align}
\label{A2-sum}
    A_2(p) &= \sum_{n=2}^\infty \sqrt{n(n-1)}
\sqrt{\binom{n+r-3}{n-2}\binom{n+r-1}{n}}\, p^{n-2}\,.
\end{align}
Since the position and momentum operators are defined as 
\begin{equation}
X = \frac{a+a^\dagger}{\sqrt{2}}, \qquad
P = \frac{a-a^\dagger}{i\sqrt{2}}\,,
\end{equation}
their expectation values in the gSG state are 
\begin{align}
    \<X\> = (1-p)^r A_1(p)\frac{z+z^*}{\sqrt{2}}~,~\qquad \<P\> = (1-p)^r A_1(p)\frac{z-z^*}{i\sqrt{2}}~,
\end{align}
To calculate the corresponding variances, we also need to compute 
\begin{align}
\langle X^2 \rangle
&=
\frac{1}{2}
\left(
\langle a^2 \rangle + \langle a^{\dagger2} \rangle
+ 2\langle N \rangle + 1
\right)~,\qquad \langle P^2 \rangle
=
-\frac{1}{2}
\left(
\langle a^2 \rangle + \langle a^{\dagger2} \rangle
- 2\langle N \rangle - 1
\right)~.
\end{align}
Plugging in Eqs.~\eqref{eq:expect-number}, \eqref{eq:expect-a} and \eqref{eq:expect-a2}, we obtain 
\begin{align}
\langle X^2 \rangle &=
\frac{1}{2}
\left[
(1-p)^r (z^2 + z^{*2}) A_2(p)
+ 2\frac{r p}{1-p} + 1
\right],
\\
\langle P^2 \rangle &=
\frac{1}{2}
\left[
-(1-p)^r (z^2 + z^{*2}) A_2(p)
+ 2\frac{r p}{1-p} + 1
\right].
\end{align}

Hence, the variances $\Delta X^2 = \<X^2\>-\<X\>^2$, $\Delta P^2 = \<P^2\>-\<P\>^2$, are 
\begin{align}
\label{varx}
\Delta X^2
=
\frac{1}{2}
+
\frac{r p}{1-p}
+
\frac{(1-p)^r}{2}(z^2+z^{*2}) A_2(p)
-
\frac{(1-p)^{2r}}{2}(z+z^*)^2 A_1(p)^2\,.
\\[10pt]
\Delta P^2
=
\frac{1}{2}
+
\frac{r p}{1-p}
-
\frac{(1-p)^r}{2}(z^2+z^{*2}) A_2(p)
+
\frac{(1-p)^{2r}}{2}(z-z^*)^2 A_1(p)^2\,.
\label{varp}
\end{align}
We see immediately that the state $|z;r\>$ is in general not a minimum certainty state. However in the $r\to \infty$ limit, we should recover the minimum certainty Glauber coherent state. We will verify this below. 
In order to take the $r\to\infty$ limit appropriately we need to rescale $z$ (recall $p$ was $|z|^2$) as
\be
z = \frac{\alpha}{\sqrt{r}}, \qquad p = \frac{|\alpha|^2}{r}~, \qquad (\alpha~\text{fixed})~.
\ee
In this limit we have the following asymptotics for various terms that appear in the variances:
\begin{align}
(1-p)^r &= e^{-|\alpha|^2}
\left(1 - \frac{|\alpha|^4}{2r} + \cdots \right), \\
A_1 &= \sqrt{r} e^{|\alpha|^2}
\left(1 + \frac{|\alpha|^4 + |\alpha|^2}{2r} + \cdots \right), \\
A_2 &= r e^{|\alpha|^2}
\left(1 + \frac{(|\alpha|^2 +1)^2}{2r} + \cdots \right).
\end{align}
Plugging these expansions in the formula for the variances in \eqref{varx} and \eqref{varp} we get:
\begin{equation}
\Delta X^2
=
\frac{1}{2}
+
\frac{1}{4r}
\left(
\alpha^2 + \alpha^{*2}
\right)
+ O(r^{-2})
\end{equation}

\begin{equation}
\Delta P^2
=
\frac{1}{2}
-
\frac{1}{4r}
\left(
\alpha^2 + \alpha^{*2}
\right)
+ O(r^{-2})
\end{equation}
So that
\begin{align}
    \Delta X^2\Delta P^2 = \frac{1}{4} + O(r^{-2})~,
\end{align}
Hence the leading deviation from the minimal uncertainty state appears at order $1/r^2$. 
\subsection{Calculation of $C(r)$ and $D(r)$}
We will provide here the details of the $p\to 1$ limit of the various formulas in the main text. 
First, we look at the sum $A_1(p)$ defined in Eq.~\eqref{A1-real-sum}. The sum converges for $p<1$ and diverges as $p\to 1$. In this regime, most of the contribution comes from the tail of the sum where $m$ is large. Using
$\binom{m+r-1}{m}\sqrt{m+r}\sim m^{r-1/2}/\Gamma(r)$, writing $p\sim e^{-\epsilon}$ and approximating the sum with an integral we get
\begin{equation}
A_{1} \approx \frac{1}{\Gamma(r)}
\int_{0}^{\infty}t^{r-\frac{1}{2}}e^{-t\epsilon}\,dt
= \frac{\Gamma\!\left(r+\tfrac{1}{2}\right)}{\Gamma(r)\,\epsilon^{r+\frac{1}{2}}}.
\label{eq:A1_asymp}
\end{equation}
Therefore,
\begin{equation}
(1-p)^{2r}p\,A_{1}(p)^{2}
\approx \epsilon^{2r}\cdot\frac{\Gamma(r+\tfrac{1}{2})^{2}}
                                 {\Gamma(r)^{2}\,\epsilon^{2r+1}}
= \frac{\Gamma(r+\tfrac{1}{2})^{2}}{\Gamma(r)^{2}\,\epsilon}.
\label{eq:A1sq_asymp}
\end{equation}

Next we look at the sum $A_2(p)$ defined in Eq.~\eqref{A2-real-sum}. Again, in the $p \to 1$ limit, most the contribution comes from the tail of the sum. At large $m$, the factor $\sqrt{(m+r)(m+r+1)}$ in the summand can be approximated as
\begin{align}
    \sqrt{(m+r)(m+r+1)} \approx m+r+\frac12 -\frac{1}{8(m+r)}~.
\end{align}
Plugging this into Eq.~\eqref{A2-real-sum}, the sum can be carried out in closed form,
\begin{align}
    A_2(p) &\approx \frac{(1-p)^{-r} (p-2 r-1)}{2 (p-1)}-\frac{\, _2F_1(r,r;r+1;p)}{8 r}~,
\end{align}
Along with the prefactor $(1-p)^r p$, we obtain the following asymptotics in small $\epsilon = 1-p$:
\begin{align}
    (1-p)^r p A_2(p) &\approx \frac{r}{\epsilon }+\left(\frac{1}{2}-r\right)-\left(\frac{1}{2}+\frac{1}{8 (r-1)}\right) \epsilon+O\left(\epsilon ^2\right)~.
\end{align}
Inserting these results into the expression of $(\Delta X)^2$ and $(\Delta P)^2$ in Eq.~\eqref{eq:dx2_real} and Eq.~\eqref{eq:dp2_real}, the leading order results are 
\begin{align}
    (\Delta X)^2 & \approx \frac{1}{\epsilon}\left(2r
   - \frac{2\,\Gamma(r+\tfrac{1}{2})^{2}}{\Gamma(r)^{2}}\right) \equiv \frac{C(r)}{\epsilon} ~,\\[5pt]
    (\Delta P)^2 & \approx \left(\frac{1}{2}+\frac{1}{8 (r-1)}\right) \epsilon \equiv D(r)\epsilon ~.
\end{align}

\bibliographystyle{JHEP.bst}
\bibliography{references}

\end{document}